\documentclass{article}

\usepackage{microtype}
\usepackage{graphicx}
\usepackage{subfigure}
\usepackage{booktabs} 

\usepackage[utf8]{inputenc} 
\usepackage[T1]{fontenc}    
\usepackage{booktabs}       
\usepackage{amsfonts}       
\usepackage{nicefrac}       
\usepackage{microtype}      
\usepackage{floatrow}
\newfloatcommand{capbtabbox}{table}[][\FBwidth]

\usepackage[table,xcdraw]{xcolor}
\usepackage{multirow}
\usepackage{graphicx}
\usepackage{verbatim} 
\usepackage{amsfonts}
\usepackage{placeins}

\usepackage[accepted,nohyperref]{icml2019}
\usepackage{float}
\floatstyle{plaintop}
\restylefloat{table}
\usepackage{titlesec}
\makeatletter
\@addtoreset{section}{part}
\makeatother
\titleformat{\part}[display]
{\LARGE\bfseries}{}{0pt}{}
\definecolor{myblue}{rgb}{0,0.08,0.45}
\usepackage{hyperref}       
\hypersetup{
    colorlinks,
    citecolor=myblue,
    filecolor=myblue,
    linkcolor=myblue,
    urlcolor=myblue
}

\icmltitlerunning{Designing antiviral candidates for SARS-CoV-2 with conditional generative models}

\begin{document}

\twocolumn[
\icmltitle{PaccMann\textsuperscript{RL} on SARS-CoV-2: Designing antiviral candidates with conditional generative models}



\icmlsetsymbol{equal}{*}
\begin{icmlauthorlist}
\icmlauthor{Jannis Born}{ibm,eth,equal}
\icmlauthor{Matteo Manica}{ibm,equal}
\icmlauthor{Joris Cadow}{ibm,equal}
\icmlauthor{Greta Markert}{ibm,eth}
\icmlauthor{Nil Adell Mill}{ibm,eth}
\icmlauthor{Modestas Filipavicius}{ibm,eth}
\icmlauthor{María Rodríguez Martínez}{ibm}
\end{icmlauthorlist}

\icmlaffiliation{ibm}{IBM Research Europe, Switzerland.}
\icmlaffiliation{eth}{ETH Zurich, Switzerland}
\icmlcorrespondingauthor{J.B.}{jab@zurich.ibm.com}
\icmlcorrespondingauthor{M.M}{tte@zurich.ibm.com}
\icmlcorrespondingauthor{J.C}{dow@zurich.ibm.com}
\icmlkeywords{Machine Learning, ICML, Conditional Generative Models, Drug Design, COVID19}
\vskip 0.3in
]



\printAffiliationsAndNotice{\icmlEqualContribution} 
\begin{abstract}
\vspace{-2mm}
With the fast development of COVID-19 into a global pandemic, scientists around the globe are desperately searching for effective antiviral therapeutic agents.
Bridging systems biology and drug discovery, we propose a deep learning framework for conditional \textit{de novo} design of antiviral candidate drugs tailored against given protein targets.
First, we train a multimodal ligand--protein binding affinity model on predicting affinities of antiviral compounds to target proteins and couple this model with pharmacological toxicity predictors.
Exploiting this multi-objective as a reward function of a conditional molecular generator (consisting of two VAEs), we showcase a framework that navigates the chemical space toward regions with more antiviral molecules.
Specifically, we explore a challenging setting of generating ligands against unseen protein targets by performing a leave-one-out-cross-validation on 41 SARS-CoV-2-related target proteins.
Using deep RL, it is demonstrated that in 35 out of 41 cases, the generation is biased towards sampling more binding ligands, with an average increase of 83\% comparing to an unbiased VAE.
We present a case-study on a potential Envelope-protein inhibitor and perform a synthetic accessibility assessment of the best generated molecules is performed that resembles a viable roadmap towards a rapid in-vitro evaluation of potential SARS-CoV-2 inhibitors.
\end{abstract}
\vspace{-4mm}
\section{Introduction}
The Severe Acute Respiratory Syndrome (SARS) Coronavirus disease (COVID 2019) is an acute respiratory disease caused by novel coronavirus SARS-CoV-2 that, to date, has infected millions and killed hundreds of thousands.
Despite longstanding efforts into understanding the pathogenicity of coronaviruses (CoVs)~\citep{drosten2003identification}, there are no approved drugs against CoV, and new systematic approaches to identify effective antiviral agents are urgently needed.
Current efforts are predominantly focused on drug repurposing strategies, with a handful of promising candidates, including
remdesivir and hydroxychloroquine.
Initial hopes are currently balked, remdesivir does not significantly reduce time to clinical improvement~\citep{wang2020remdesivir2} and hydroxychloroquine was not found effective in a meta-study of human clinical trials~\citep{Shamshirian2020Hydroxychloroquine}.
\citet{gordon2020sars} recently identified 69 promising compounds by measuring binding affinities of 26 out of the 29 SARS-CoV-2 proteins against human proteins. 

\begin{figure}[htb!]
\centering
\includegraphics[width=.99\columnwidth]{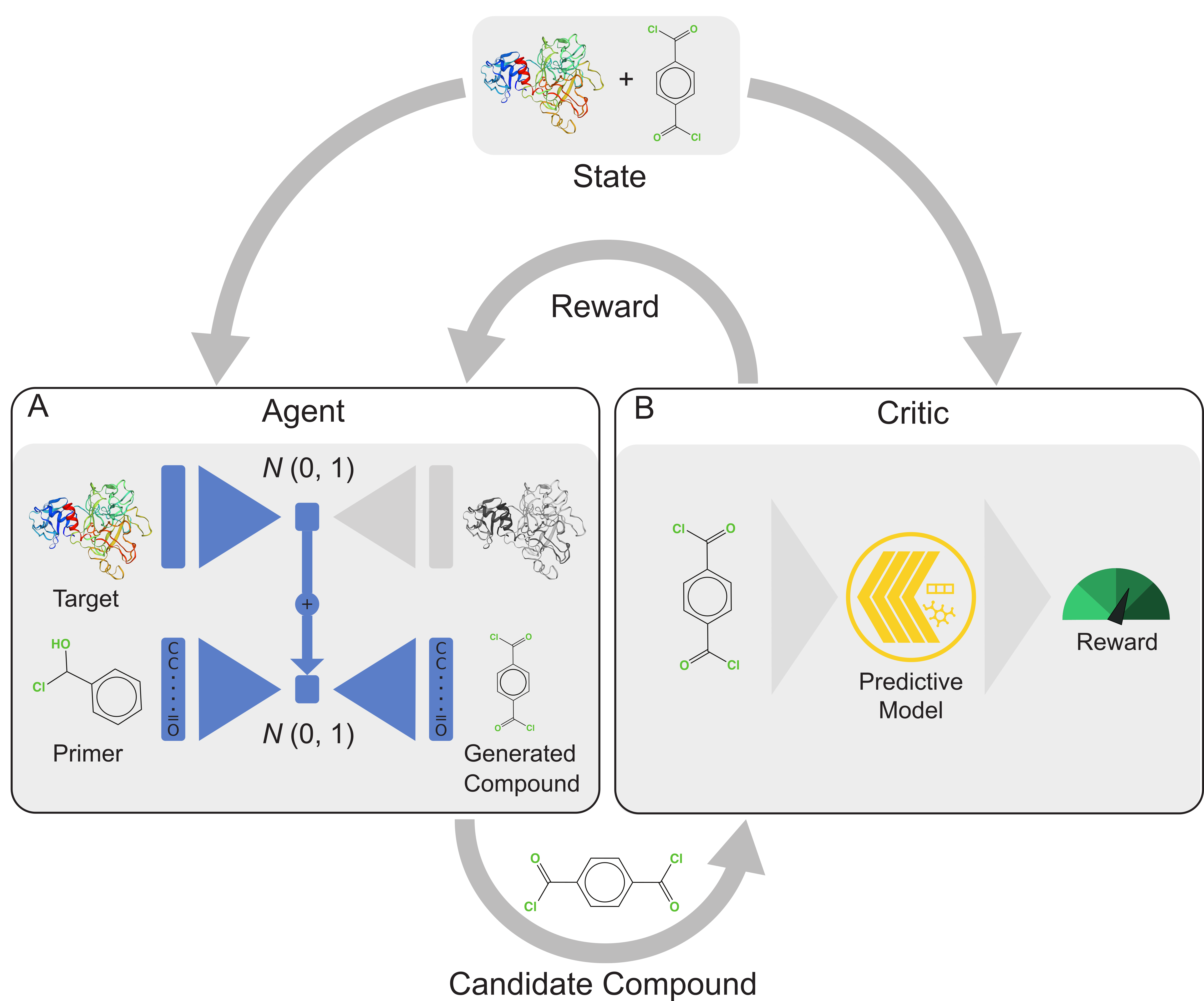}
\caption{\textbf{A drug discovery framework for antiviral small molecules against SARS-CoV-2.} 
The conditional compound generator, called agent (see \textbf{A}), can produce novel structures specifically designed to target a protein of interest.
The generative process starts with the encoding of the primary structure of the target protein into a latent space of protein sequences. The representation is fed into a molecular decoder of a separately pretrained molecule VAE to produce a candidate compound.
Next, the proposed compound is evaluated by a critic (see \textbf{B}) composed by: a multimodal deep learning model that predicts protein-drug binding affinity using protein and compound sequences as input, and a QSAR-based score to punish toxicity.
By means of the reward given by the critic, a closed-loop system is created and is trained with deep reinforcement learning to maximize a multi-objective reward.}
\label{fig:model}
\end{figure}
With high uncertainty in the outcome of drug repurposing strategies, it is worth exploiting \textit{de novo} drug discovery approaches against SARS-CoV-2.
Drug discovery is challenging, with costs of up to 3 billion US\$ per new FDA-approved drug, an attrition rate of 99.99\%, more than 10 years until market release and a search space of $10^{60}$ compounds~\citep{scannell2012diagnosing}. 
However, the availability of high-throughput screenings of compound--protein interactions (CPI) has enabled deep learning to set new benchmarks for large-scale QSAR prediction models for predicting protein--drug binding affinity~\citep{karimi2019deepaffinity}.
Deep learning has further been proven feasible of \textit{in silico} design of molecules with desired chemical properties and shown potential to accelerate discovery of DDR1 inhibitors~\citep{zhavoronkov2019deep}. 
A few studies used deep generative models to release libraries of (unsynthesized) candidates to target 3C-like protease, a main therapeutic target of SARS-CoV-2~\citep{zhavoronkov2020potential,tang2020ai} but both studies manually curated datasets to target 3C-like protease inhibitors.
Here, we aim to bridge systems biology and drug discovery, using deep learning to explore target-driven drug design with conditional generative models.
Our framework (see~\autoref{fig:model}) for conditional molecular design is conceptually inspired by our previous work,~\textbf{PaccMann\textsuperscript{RL}}~\citep{born2020paccmannrl}, however note that here we focus on protein-driven instead of omics-profile-driven drug generation.
Our framework can be trained to design compounds against any primary protein structure(s).
Deep learning for target-driven drug design was first formulated by~\citet{aumentado2018latent} and similarly to~\citet{chenthamarakshan2020target}, our approach implements a conditional generator that can be applied to \textit{unseen} protein targets. However, instead of using conditional sampling, we perform an RL-biased conditional generation (fusing the latent spaces of protein targets and small molecules) that is demonstrated to generalize to \textit{unseen} targets.
We further couple our model with \texttt{IBM RXN}\footnote{\url{https://rxn.res.ibm.com/}}, an AI-governed platform for automated chemical synthesis to promptly synthesize the best compounds~\citep{schwaller2020predicting}. 

\section{Methods}
\textbf{SELFIES VAE.}
The molecular generator is a variational auto-encoder (VAE)) that is pretrained on 1,576,904 bioactive compounds from \href{https://www.ebi.ac.uk/chembl/}{ChEMBL} (10\% are held out as validation set).
The VAE implementation mostly follows~\citet{born2020paccmannrl}, i.e., it consists of two layers of stack-augmented GRUs~\citep{joulin2015inferring} in both encoder and decoder.
The latent space has a dimensionality of 256, molecules are represented as SELFIES~\citep{krenn2019selfies}, a robust adaption of the molecular in-line notation SMILES (Simplified Molecular Input Line Entry Specification) devised for generative models, and one-hot encodings are used.
During training, KL annealing, teacher forcing and token dropout are employed. During testing, the stochastic decoder is sampling from the softmax distribution over the output tokens.

\textbf{Protein VAE.}
The protein VAE consists of 3 dense layers of sizes [768, 512, 256] in both encoder and decoder. The model is trained on $\sim$400,000 proteins from \href{https://www.uniprot.org}{UniProt} (SwissProt).
The proteins considered were selected by filtering out sequences longer than 8,190 amino acids.
The maximum length of the sequences has been selected to accommodate the statistics for the SARS-CoV-2 relevant proteins compiled by UniProt\footnote{\url{covid-19.uniprot.org/}as on 22 May 2020.}
.
Note that the VAE is not trained on the raw sequences but on 768 dimensional latent representations obtained from TAPE~\citep{rao2019evaluating}.
During trainining, KL annealing and dropout are employed.


\textbf{Protein-ligand affinity prediction.}
\label{p:predictive}
%
%
To predict CPI, we utilize a bimodal neural network based on the multiscale convolutional attention model~\citep{manica2019toward, cadow2020paccmann} (MCA, for model architecture see appendix~\autoref{fig:predictor}).
Drug--protein binding affinity data is obtained from \href{https://www.bindingdb.org/bind/index.jsp}{BindingDB}, a public database of 1,813,527 measured binding affinities between 7,044 proteins and 802,551 small drug-like compounds as of March 2020. 
As for the protein VAE data, the database was filtered from entries with target sequences longer than 8,190 amino acids.
The remaining 1,361,076 entries 
with an average of 187 reported compounds per target protein were taken as binding examples.
From compounds not reported as entry for a given target 187 compounds were randomly sampled as non--binding to the respective protein target to match the binding examples. Finally, the combined examples were filtered for invalid SMILES to a total of 2,723,726 binding/non--binding pairs of 771,839 compounds and split into random, stratified train (72\%), validation (18\%) and test (10\%) folds.

\textbf{Toxicity prediction.}
Using the Tox21 database available through DeepChem~\citep{wu2018moleculenet}, we trained a MCA model on the augmented SMILES sequences~\citep{bjerrum2017smiles} to predict the 12 toxicity classes.

\textbf{Conditional generation.}
The conditional generative model is obtained by encoding a protein target with the protein VAE and decoding it with the pretrained molecular decoder.

The objective function of this hybrid VAE $G_{\Theta}$ is:
\begin{equation}
    \Pi(\Theta|r) = \sum_{s_T \in S^*} P_{\Theta}(s_T|r)R(s_T,r)
    \label{eq:policy_gradient}
\end{equation}
where $r$ is the protein target of interest, $s_T$ is a SELFIES string at time $T$ (terminated with the \texttt{<END>} token), $S^{*}$ resembles the molecular space and $P_{\Theta}(s_T|r)$ is the probability to sample $s_T$ given $r$. 
In detail, $P_{\Theta}(s_T|r):= \prod_{t=0:T}p(a_t|s_{t-1})$
where $s_0=r$ and $a_t$ is the action at time $t$ sampled from the dictionary of SELFIES tokens.
$R(s_T,r)$ is the output of the critic $C$, in our case a multi objective:
\begin{equation}
    R(s_T,r) = A(s_T, r) +0.5\; T(s_T)
    \label{eq:reward}
\end{equation}
where $A(\cdot)$ is the affinity predictor and $T(\cdot)$ is the toxicity predictor that returns 1 iff $s_T$ is inactive in all 12 assays.
Since~\autoref{eq:policy_gradient} is intractable to compute, it is approximated using policy gradient and subject to maximization using REINFORCE~\citep{williams1992simple}.

\textbf{Protein targets.}
We fetched the 41 protein targets that are labelled as relevant to SARS-CoV-2 by UniProt (as on 22 May 2020).
A full list of targets is given in~\autoref{tab:results} and includes e.g. the 3C-like protease (M\textsubscript{pro}) which was identified as most promising candidate for antiviral compound development~\citep{wu2020analysis} and was already investigated with generative models~\citep{zhavoronkov2020potential} and molecular docking studies~\citep{khaerunnisa2020potential}. 
Other proteins are the nucleocapsid (N-) protein or the spike glycoprotein which is the most important surface protein, the target of chloroquine and mediates entrance to human respiratory epithelial cells by interacting with the ACE2 receptor. 

\section{Results}
\textbf{Protein-ligand affinity prediction.}
Because the conditional generation focuses on antiviral drug design, it is important that the affinity predictor generalizes well for viral proteins. 
The results of the MCA model on validation and test data from BindingDB are displayed in~\autoref{tab:predictor} next to the performance on ~10k viral proteins.
\begin{table}[!htb]
\centering
\begin{tabular}{llllll}
\toprule
& \textbf{Validation} &   \textbf{Test} & \textbf{Viral}
\\ \midrule
\textbf{ROC-AUC} & 0.968 & 0.969 & 0.96 \\
\textbf{Average precision} & 0.963 & 0.965 & 0.92 \\
\bottomrule          
\end{tabular}

\caption{Result of bimodal affinity predictor on BindingDB data.}
\label{tab:predictor}
\end{table}
The results shows that the model learned reasonably well to classify CPI as binding or non-binding.
         
\textbf{Toxicity predictor.}
Because toxicity is a major cause of the high attrition rate in drug discovery, we decided to perform a multi-objective optimization (see ~\autoref{eq:reward}) based on toxicity and binding affinity.
Across 10 runs, this model achieved a ROC-AUC of 0.877 $\pm$ 0.04, surpassing prior results on this benchmarked dataset.
Both the affinity and toxicity predictor are not investigated further herein, but employed as reward function for the conditional generation.

\textbf{Conditional generative model\newline}
\label{generative-model}
In this study, we are not primarily interested in proposing the \textit{best} possible compounds.
We rather want to validate whether our framework can go beyond current approaches for target-driven compound design~\citep{chenthamarakshan2020target,zhavoronkov2019deep,zhavoronkov2020potential} in the sense that it does not require fine-tuning for specific targets.
We therefore investigated the generalization capabilities of our framework by performing a leave-one-out-cross-validation (LooCV) on the 41 targets. The RL optimizaton was performed for 5 epochs and 500 molecules were sampled in each step. 
The results are depicted in~\autoref{tab:results} and demonstrate that in 35 out of 41 cases the model proposed more binding compounds against an unseen target, compared to the baseline SELFIES VAE.

\begin{table}[!htb]
\centering
\scalebox{0.6}{
\begin{tabular}{llllll}
\toprule
          \textbf{Target protein} &       \textbf{Affinity\textsubscript{0}} &      \textbf{Aff\textsubscript{med}$\pm$SEM} & \textbf{Aff\textsubscript{b}} &      \textbf{Tox\textsubscript{med}$\pm$SED} & \textbf{Tox\textsubscript{b}} \\
\midrule
       VME1-CVHSA &      20\% &   18\% $\pm$ 3\% &     \textbf{29\%} &   6\% $\pm$ 3\% &     19\% \\
       IMA1-HUMAN &      88\% &   97\% $\pm$ 1\% &    \textbf{100\%} &   5\% $\pm$ 3\% &     18\% \\
             VEMP-SARS2 &       \textbf{29\%} &   16\% $\pm$ 2\% &    20\% &   9\% $\pm$ 2\% &      12\% \\
       NS7B-SARS2 &      25\% &   30\% $\pm$ 5\% &    \textbf{ 33\%} &   7\% $\pm$ 5\% &     25\% \\
       ITAL-HUMAN &      24\% &   16\% $\pm$ 6\% &    \textbf{ 43\%} &   9\% $\pm$ 1\% &     12\% \\
       NCAP-CVHSA &      \textbf{17\%} &   11\% $\pm$ 1\% &     15\% &  12\% $\pm$ 2\% &     14\% \\
       R1AB-CVHSA &      58\% &   90\% $\pm$ 2\% &    \textbf{ 91\%} &   9\% $\pm$ 1\% &     11\% \\
       NS8B-CVHSA &       9\% &   12\% $\pm$ 2\% &    \textbf{ 20\%} &   7\% $\pm$ 4\% &     25\% \\
 A0A663DJA2-SARS2 &      26\% &   35\% $\pm$ 3\% &    \textbf{ 41\%} &  14\% $\pm$ 3\% &     18\% \\
       NS8A-CVHSA &      21\% &   47\% $\pm$ 4\% &    \textbf{ 55\%} &  10\% $\pm$ 1\% &     10\% \\
       NS7A-SARS2 &       4\% &    3\% $\pm$ 1\% &    \textbf{  7\%} &  10\% $\pm$ 3\% &     19\% \\
        Y14-SARS2 &      17\% &   29\% $\pm$ 4\% &    \textbf{ 43\%} &   8\% $\pm$ 2\% &     14\% \\
        NS6-SARS2 &      20\% &   12\% $\pm$ 3\% &    \textbf{ 22\%} &   4\% $\pm$ 3\% &     14\% \\
      SMAD3-HUMAN &      50\% &   74\% $\pm$ 3\% &    \textbf{ 86\%} &   6\% $\pm$ 1\% &     10\% \\
      SPIKE-CVHSA &       3\% &    0\% $\pm$ 1\% &    \textbf{  5\%} &   7\% $\pm$ 1\% &     11\% \\
       DDX1-HUMAN &       9\% &   14\% $\pm$ 2\% &    \textbf{ 20\%} &   9\% $\pm$ 1\% &     10\% \\
       AP3A-SARS2 &       \textbf{4\%} &    0\% $\pm$ 1\% &      3\% &   9\% $\pm$ 3\% &     19\% \\
        R1A-CVHSA &      14\% &   45\% $\pm$ 3\% &    \textbf{ 50\%} &   9\% $\pm$ 1\% &     11\% \\
        NS8-SARS2 &       7\% &   10\% $\pm$ 3\% &    \textbf{ 18\%} &  10\% $\pm$ 1\% &     15\% \\
       PHB2-HUMAN &       4\% &    3\% $\pm$ 0\% &    \textbf{  4\%} &  11\% $\pm$ 3\% &     23\% \\
       SGTA-HUMAN &      11\% &   12\% $\pm$ 1\% &    \textbf{ 13\%} &   8\% $\pm$ 1\% &     12\% \\
       NS7A-CVHSA &      18\% &   35\% $\pm$ 5\% &    \textbf{ 59\%} &  11\% $\pm$ 2\% &     15\% \\
      ORF9B-CVHSA &       9\% &   11\% $\pm$ 2\% &    \textbf{ 17\%} &   6\% $\pm$ 1\% &     11\% \\
        R1A-SARS2 &      62\% &   82\% $\pm$ 3\% &    \textbf{ 89\%} &   8\% $\pm$ 2\% &     14\% \\
        Y14-CVHSA &      14\% &   15\% $\pm$ 2\% &    \textbf{ 23\%} &  11\% $\pm$ 2\% &     15\% \\
      ORF9B-SARS2 &      \textbf{18\%} &   12\% $\pm$ 1\% &     15\% &  12\% $\pm$ 2\% &     16\% \\
      TMPS2-HUMAN &       6\% &    5\% $\pm$ 1\% &    \textbf{  6\%} &   6\% $\pm$ 1\% &     10\% \\
       BST2-HUMAN &      10\% &    5\% $\pm$ 3\% &    \textbf{ 16\%} &  10\% $\pm$ 2\% &     14\% \\
       NS3B-CVHSA &      25\% &   23\% $\pm$ 2\% &    \textbf{ 29\%} &  12\% $\pm$ 1\% &     15\% \\
      SPIKE-SARS2 &       7\% &    6\% $\pm$ 2\% &    \textbf{ 12\%} &  10\% $\pm$ 1\% &     12\% \\
      FURIN-HUMAN &      28\% &   27\% $\pm$ 4\% &    \textbf{ 36\%} &   9\% $\pm$ 3\% &     20\% \\
       AP3A-CVHSA &       \textbf{9\%} &    0\% $\pm$ 1\% &      6\% &   8\% $\pm$ 1\% &     12\% \\
       VME1-SARS2 &      15\% &   16\% $\pm$ 3\% &    \textbf{ 27\%} &   6\% $\pm$ 2\% &     14\% \\
       NS7B-CVHSA &      21\% &   26\% $\pm$ 1\% &    \textbf{ 27\%} &   7\% $\pm$ 1\% &     11\% \\
       MPP5-HUMAN &       5\% &    9\% $\pm$ 2\% &    \textbf{ 11\%} &  15\% $\pm$ 2\% &     16\% \\
       ACE2-HUMAN &      51\% &   77\% $\pm$ 4\% &    \textbf{ 85\%} &   5\% $\pm$ 2\% &     12\% \\
       VEMP-CVHSA &      21\% &   25\% $\pm$ 3\% &    \textbf{ 30\%} &  12\% $\pm$ 2\% &     20\% \\
        NS6-CVHSA &      10\% &   13\% $\pm$ 1\% &    \textbf{ 15\%} &   3\% $\pm$ 3\% &     14\% \\
        PHB-HUMAN &       3\% &    0\% $\pm$ 1\% &    \textbf{  3\%} &   6\% $\pm$ 1\% &      7\% \\
       R1AB-SARS2 &      83\% &  100\% $\pm$ 0\% &    \textbf{100\%} &   5\% $\pm$ 1\% &      7\% \\
       NCAP-SARS2 &      \textbf{25\%} &    5\% $\pm$ 2\% &      9\% &   9\% $\pm$ 4\% &     24\% \\
              \bottomrule
       \textbf{Average} &  \textbf{18\%} &     \textbf{26\% $\pm$ 4\%} &       \textbf{33\%} &    \textbf{9\% $\pm$ 0.5\%} &      \textbf{15\%} \\
       \bottomrule
\end{tabular}
}
\caption{\textbf{Generating antiviral compounds against unseen SARS-CoV-2 targets.} For each of the 41 targets, Affinity\textsubscript{0} shows the fraction of binding molecules sampled \textit{before} training.
Aff\textsubscript{best} shows the fraction at the best epoch of RL training, while Aff\textsubscript{median} shows the median across all 5 training epochs. The same applies to Tox\textsubscript{best} and  Tox\textsubscript{med}, where Tox\textsubscript{0} was 8.7\%.}
\label{tab:results}
\end{table}

From the baseline SELFIES VAE, a total of 3,000 molecules was sampled.
The average ratio of compounds predicted to bind increased from 18\% to 26\% with the best epoch averaging 33\%. For example density plots see the appendix (\autoref{fig:densities}).
We additionally optimized the generator to propose less toxic compounds.
This succeeded to a lesser extent, probably at least partially caused by the lower weight in the reward function.
\begin{figure}[htb!]
\centering
\includegraphics[width=1.\columnwidth]{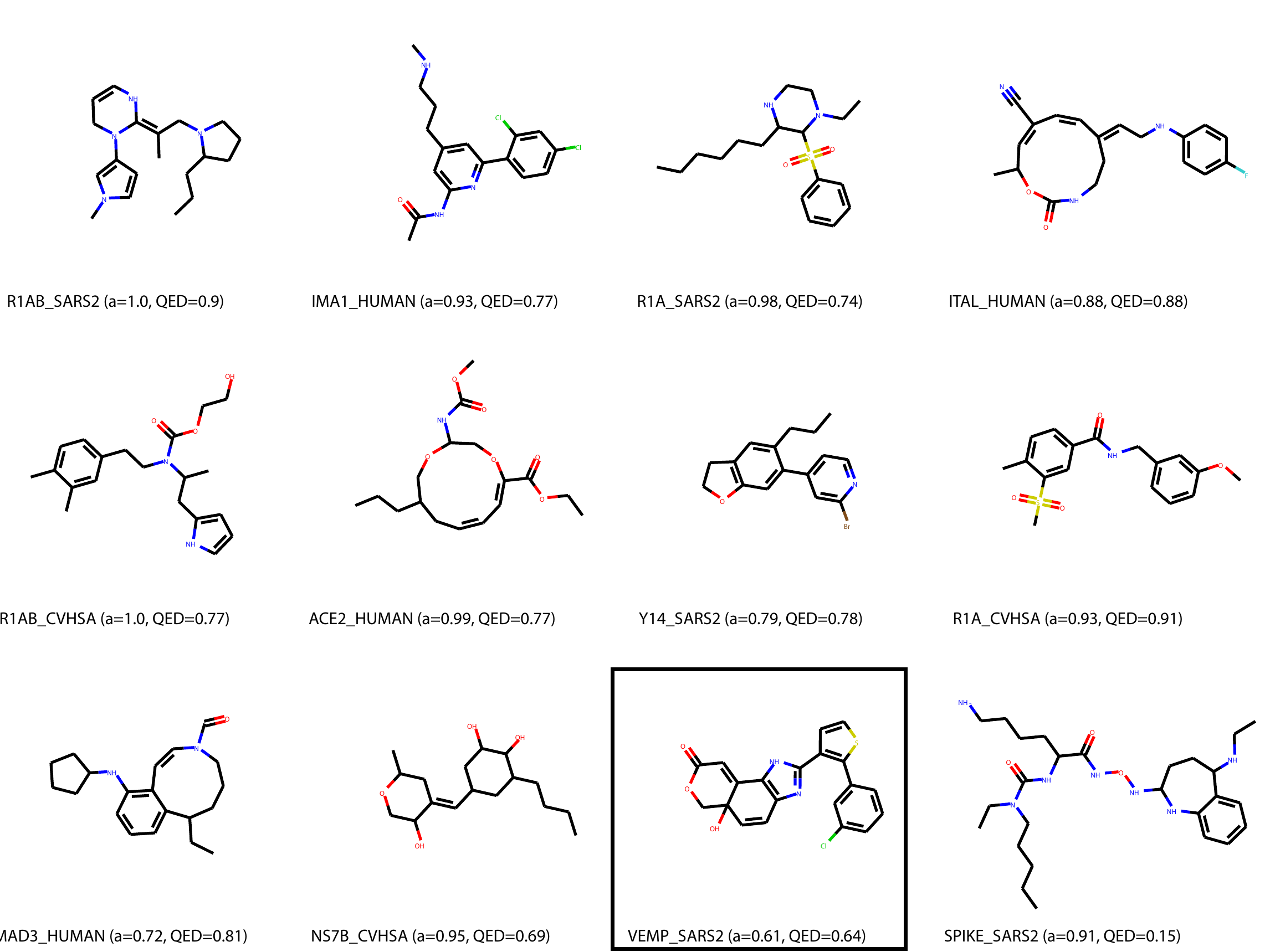}
\caption{\textbf{Molecules sampled against specific protein targets.} For a selection of 12 targets, the generated compound with the highest reward is depicted. $a$ stands for binding affinity. The encircled molecule is discussed further in the case study. 
}
\label{fig:samples}
\vspace{-2mm}
\end{figure}
For a qualitative evaluation, \autoref{fig:samples} shows a selection of the sampled molecules alongside their QED score~\citep{bickerton2012quantifying}).

To investigate the learned chemical space, we assembled a dataset of 10,000 random ChEMBL compounds, 3,000 molecules sampled from the unbiased VAE, 3,000 molecules sampled from the biased generator and 82 SARS-CoV-2 candidate drugs from the literature (top 15 matches on PubChem and 69 compounds identified via protein-interaction-maps~\citep{gordon2020sars}, excluding 2 duplicates).
For all these molecules, binding affinities were computed alongside other pharmacological properties.
Next, a UMAP~\citep{mcinnes2018umap} was performed on the ECFP4 fingerprint~\citep{rogers2010extended} and visualized with Faerun/Tmap~\citep{probst2018fun,probst2020visualization}\footnote{The \texttt{Faerun} visualization with ECFP is available at:\newline\tiny{\url{https://paccmann.github.io/assets/umap\_fingerprints.html}}}
.
The interactive visualisation shows that the RL optimisation lead to over-sampling a manifold of the chemical space that is more densely populated with binding compounds.
The 3D UMAP shows that the currently investigated candidate molecules (red) are structurally fairly dissimilar (i.e. scattered across the chemical space). 
But it gives evidence that our model successfully navigates the chemical space towards regions of high reward.
While this shows that the generator succeeded in its objective, ultimately, the quality of the reward function remains the bottleneck of the framework.

\textbf{Case study.} For a more detailed assessment of the quality of the molecules, we ranked all $\sim$ 3,000 conditionally generated molecules by their tanimoto similarity $\tau$ to the closest neighbour of the 81 literature candidates. 
Among the top 5 molecules, we found the molecule encircled in~\autoref{fig:samples},  generated against \texttt{VEMP\textsubscript{SARS2}} (UniProt ID:~\texttt{P0DTC4}), the envelope small membrane protein (E-Protein), a key player for virion assembly and morphogenesis.
Our candidate exhibits the highest tanimoto similarity to the compounds MZ1 and dBET6 ($\tau=0.64$ based on \texttt{RDKit} fingerprint).
These two pre-clinical SARS-CoV-2 drug candidates were identified by~\citet{gordon2020sars} to target the E-protein by degrading the human BRD2 and BRD4 proteins and thus preventing the virus from inducing changes in the host's protein expression.

\begin{figure}[htb!]
\centering
\includegraphics[width=1.\columnwidth]{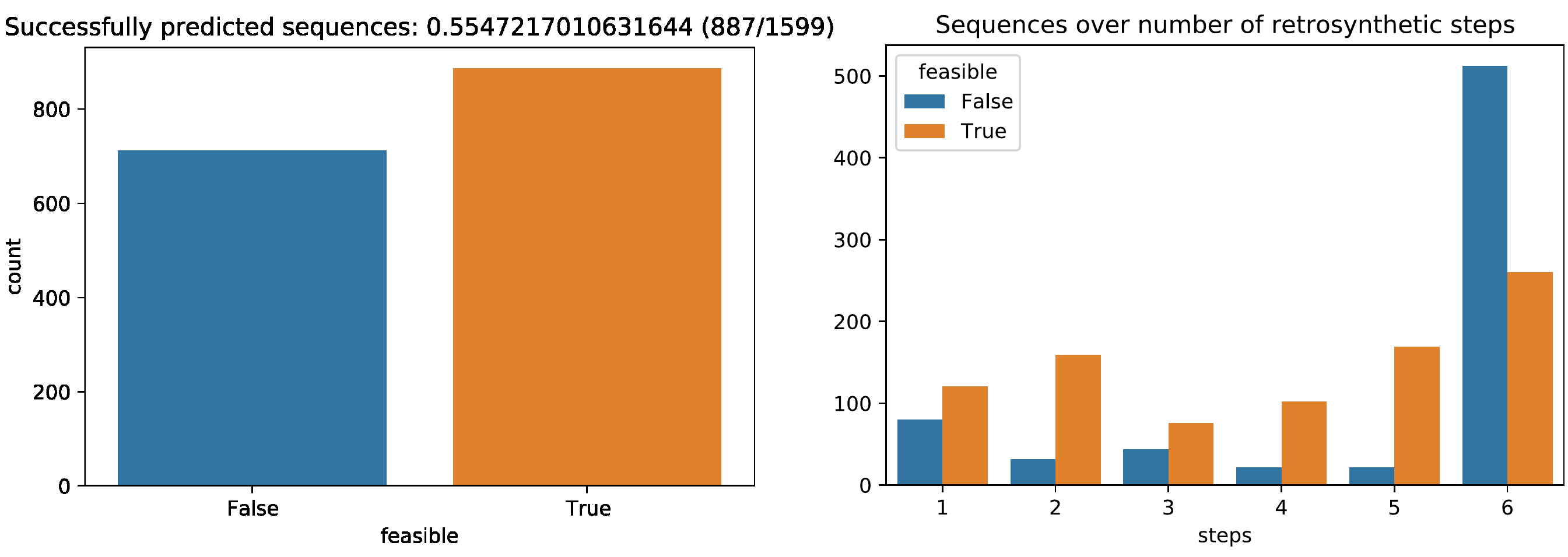}
\vspace{-8mm}
\caption{\textbf{Results of retrosynthesis attempts.}
A retrosynthetic pathway is considered feasible if it leads to commercially available precursors within six reaction steps. Orange indicates feasiblity while blue indicates non feasibility.
On the left (\textbf{A}), overall feasibility of the predicted sequences. On the right (\textbf{B}), feasibility over the number of reaction steps.
}
\label{fig:retro}
\end{figure}
The Top-5 candidate compounds for each protein target were further analyzed for synthetic feasibility using \texttt{IBM RXN}'s retrosynthesis engine~\citep{schwaller2020predicting}. We performed the predictions using the \texttt{Python} package \texttt{rxn4chemistry}
\footnote{\scriptsize{\url{https://github.com/rxn4chemistry/rxn4chemistry}}}
.
~\autoref{fig:retro} shows the predictions over the retrosynthetic sequences estimated for all molecules.
Although the generated molecules are not optimized for synthetic accessibility, more than half of the synthesis routes predicted are feasible. It's interesting to observe how more than 300 sequences requires only a single or two steps reactions, indicating, assuming a reasonable yield, an extremely efficient synthesis for part of the molecules generated.

\section{Discussion}
%


%
The dramatic effect of the COVID-19 pandemic is compounded by the lack of vaccines and therapeutic agents against SARS-CoV-2. Worse, traditional approaches to drug discovery are slow, inefficient, error-prone and costly. 


%
Here, we proposed a novel framework for compound design that can be targeted towards \emph{any viral target protein} with no retraining requirements.
We showcased the potential of the framework by successfully tackling the problem of generating novel compounds with high binding affinity to unseen targets, while controlling toxicity of the generated molecules.
Furthermore, for each target, we estimated retrosynthetic pathways of the most promising molecules, to assess the feasibility of the generated compounds.
Large-scale screening data for 1,670 compounds tested against SARS-CoV-2 proteins that just became available~\citep{heiser2020identification}, will be used in the future to improve the affinity predictor, one of the bottlenecks of our approach.

\newpage
\bibliographystyle{icml2019}
\interlinepenalty=10000
{\footnotesize
\bibliography{references}
}
\clearpage
\section*{Appendix}
\begin{figure}[htb!]
\centering
\includegraphics[width=1.\columnwidth]{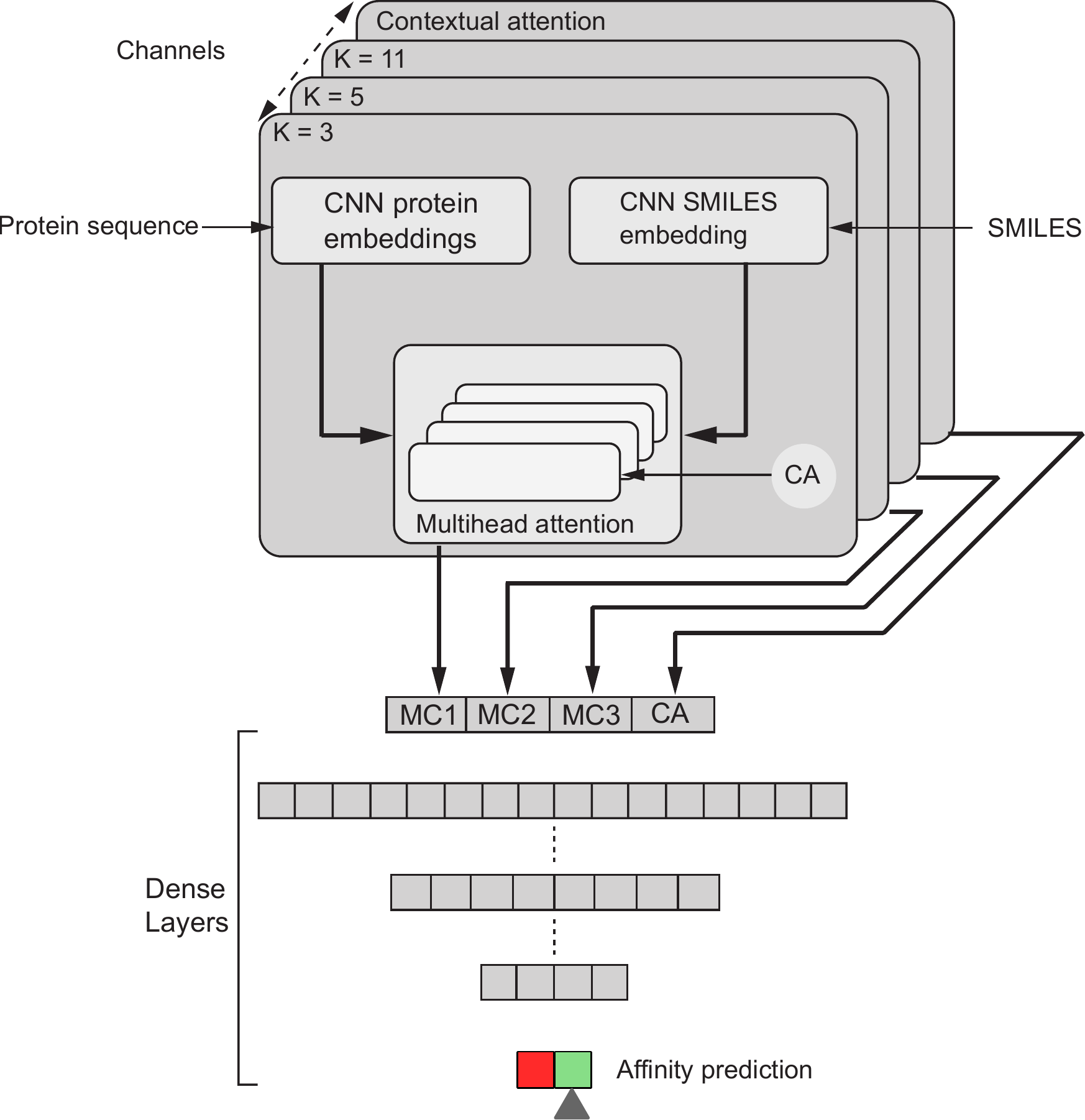}
\caption{\textbf{Multimodal protein-ligand affinity predictor for antiviral compounds.}
Inspired by the MCA architecture in~\citet{manica2019toward}, this is a multimodal classification model that performs multiscale convolutions on SMILES embeddings (ligand) and amino acid embeddngs (protein). 
The output is fed into multiple heads of contextual attention mechanism prior to a set of stacked dense layers.
}
\label{fig:predictor}
\end{figure}

\begin{figure}[htb!]
\centering
\includegraphics[width=1.05\columnwidth]{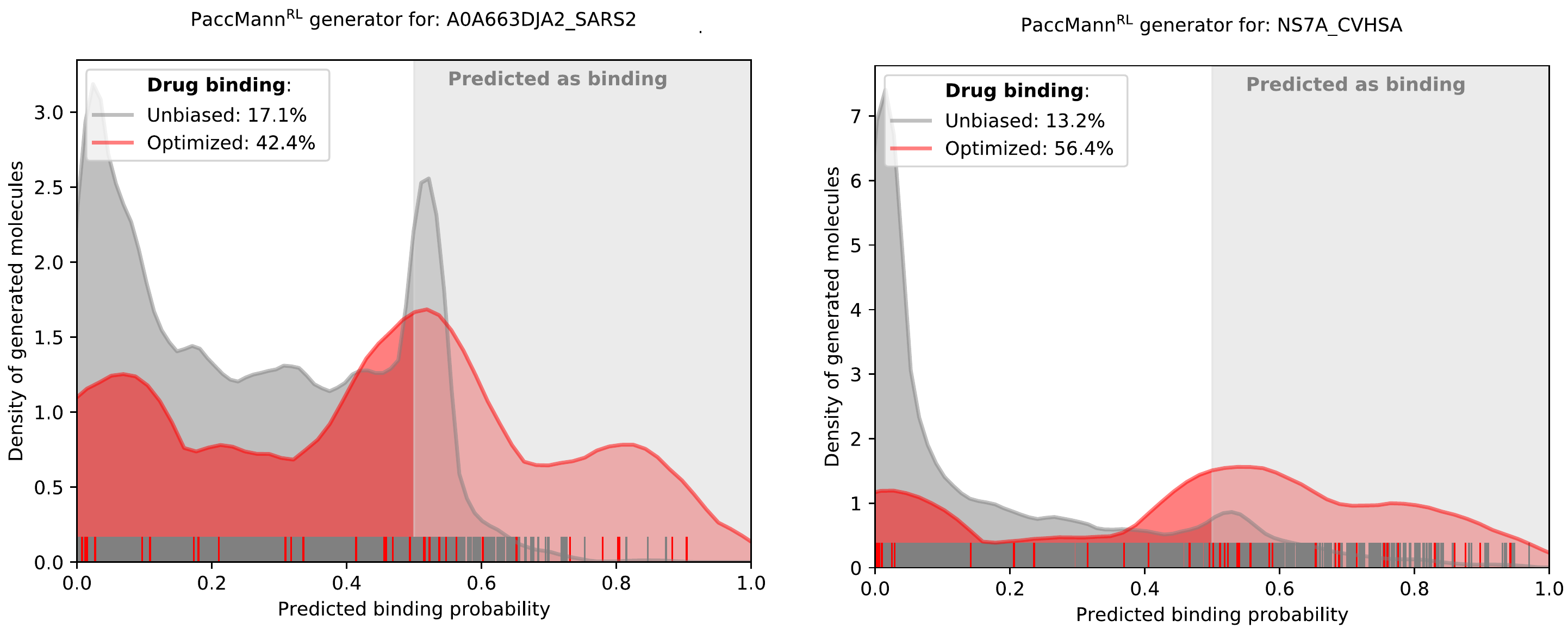}
\caption{\textbf{Exemplary density functions of conditional generation.} Gray distributions show predicted binding affinities of $n$=3,000 molecules sampled from an unbiased SELFIES VAE. Depicted in red are the densities obtained by sampling from the RL optimized conditional generative model.
It can be seen that the optimization biased the sampling process toward regions of the chemical space that are more densely populated with ligands that are predicted to bind.
}
\label{fig:densities}
\end{figure}

\begin{figure}[htb!]
\centering
\includegraphics[width=1.05\columnwidth]{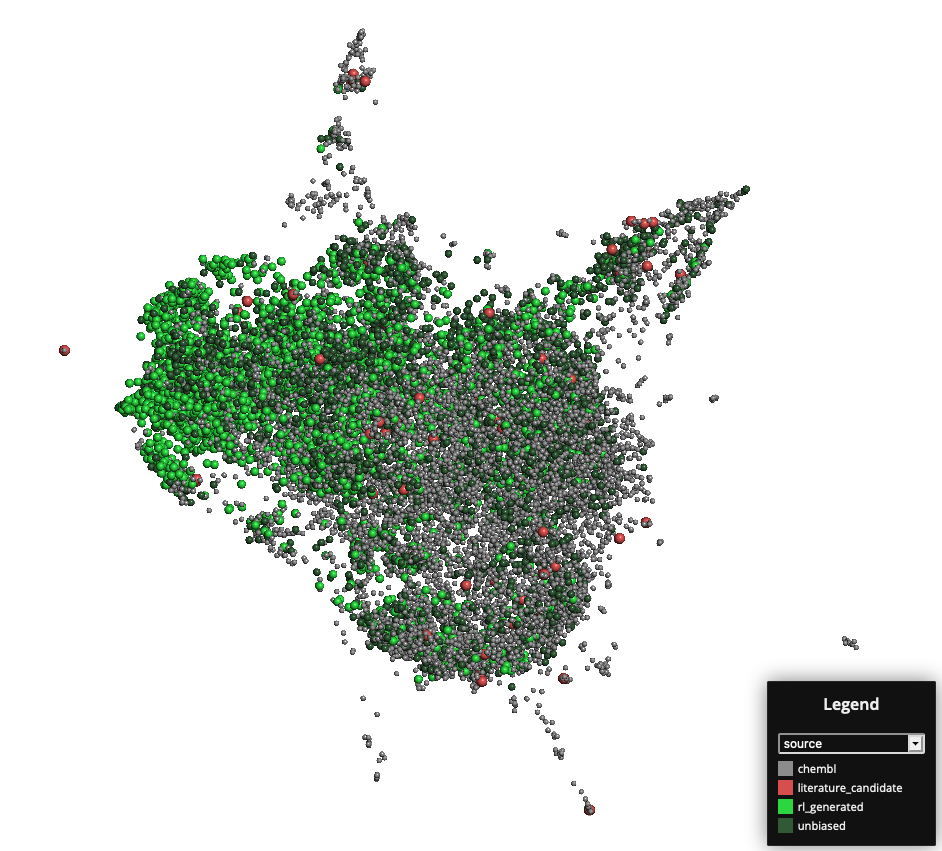}
\caption{\textbf{UMAP dimensionality reduction of the chemical space visualized with~\texttt{Faerun}/\texttt{TMap}~\citep{probst2018fun,probst2020visualization}.
}
Snapshot of the~\texttt{Faerun} visualization of a UMAP of 10,000 molecules randomly selected from ChEMBL (grey), alongside 3,000. molecules sampled from the unbiased generator (dark green), 3,500 molecules sampled against SARS-CoV-2 related target proteins and 82 drug candidates according to the literature.
}
\label{fig:chemicalspace}
\end{figure}

\end{document}